\documentclass[acmsmall,screen]{acmart}
\setcopyright{none}

\usepackage[T1]{fontenc}
\usepackage[utf8]{inputenc}
\usepackage{booktabs}
\usepackage{longtable}
\usepackage{array}
\usepackage{xcolor}
\usepackage{enumitem}
\usepackage{graphicx}
\usepackage{amsmath}
\usepackage{pgfplots}
\pgfplotsset{compat=1.17}
\usepackage{multirow}
\hypersetup{
    colorlinks=true,
    linkcolor=blue,
    citecolor=green!50!black,
    urlcolor=blue!70!black
}
\usepackage{tikz}
\usetikzlibrary{positioning,arrows.meta,calc}
\usepackage{pifont}

\usepackage[section]{placeins}
\makeatletter
\AddToHook{begindocument/end}{%
  \expandafter\renewcommand\expandafter\subsection\expandafter
    {\expandafter\@fb@secFB\subsection}%
}
\makeatother

\newcommand{\find}[1]{%
  \par\addvspace{0.75em}%
  \noindent\fbox{%
    \begin{minipage}{\dimexpr\linewidth-2\fboxsep-2\fboxrule\relax}
    \raggedright #1
    \end{minipage}}%
  \par\addvspace{0.75em}%
}

\renewcommand\footnotetextcopyrightpermission[1]{}

\title{A Low Similarity Score Does Not Show Independence after Source-Conditioned Repository Rewriting}

\author{Micheline Bénédicte Moumoula}
\orcid{0009-0005-1206-2270}
\authornote{Corresponding author (micheline.moumoula@uni.lu)}
\affiliation{%
  \institution{University of Luxembourg }
  \city{Luxembourg}
  \country{Luxembourg}
}

\email{micheline.moumoula@uni.lu}

\author{Serge Lionel Nikiema}
\orcid{0009-0001-0066-3694}
\affiliation{%
  \institution{University of Luxembourg}
  \city{Luxembourg}
  \country{Luxembourg}
}
\email{lionel.nikiema@uni.lu}

\author{Tailia Malloy}
\affiliation{%
  \institution{University of Luxembourg}
  \city{Luxembourg}
  \country{Luxembourg}
}
\email{tailia.malloy@uni.lu}

\author{Abdoul Kader Kaboré}
\orcid{0000-0002-3151-9433}
\affiliation{%
  \institution{University of Luxembourg}
  \city{Luxembourg}
  \country{Luxembourg}
}
\email{abdoulkader.kabore@uni.lu}

\author{Jacques Klein}
\orcid{0000-0003-4052-475X}
\affiliation{%
  \institution{University of Luxembourg}
  \city{Luxembourg}
  \country{Luxembourg}
}
\email{jacques.klein@uni.lu}

\author{Tegawendé F. Bissyandé}
\orcid{0000-0001-7270-9869}
\affiliation{%
  \institution{University of Luxembourg}
  \city{Luxembourg}
  \country{Luxembourg}
}
\email{tegawende.bissyande@uni.lu}

\renewcommand\shortauthors{Moumoula et al.}

\begin{abstract}
Maintainers sometimes cite a low token-similarity score as evidence that an LLM-assisted rewrite of a copyleft project is independent of the original. We test what such a score can support under \emph{source-conditioned} rewriting: a coding agent keeps the original in its working copy, rewrites
the repository, and iterates until a fixed behavioural harness passes, with no human edit to the code. On 69 rewrites of six copyleft Python projects we find that: \ding{182}~Agent-only rewriting is inexpensive and often yields low scores. Every controlled rewrite passes the harness after repair, usually within half an hour for the initial rewrite. Under a ground-up rearchitecture prompt, most rewrites fall below 25\% JPlag similarity and still pass the harness.
\ding{183}~A single low score is an unreliable witness. Identical project--agent--prompt configurations produce wide spreads in similarity; run-to-run variation dominates project and agent effects, and in more than a third of configurations the three repeats straddle the
threshold. Rewrites retain more of the original's imports and fingerprint facts than unrelated projects do, yet hand-written independent implementations of the same task often retain as much, and neither signal separates the two groups better than chance. Thus a low score on one rewrite is weak evidence of independence: another
run of the same agent may score high. Claims based on similarity should report repeated generations, and candidate provenance signals should be checked against same task independent implementations, not only against unrelated projects.
\end{abstract}



\begin{document}
\maketitle

\section{Introduction}
\label{sec:introduction}

In March 2026, version 7.0.0 of \texttt{chardet}, a widely used Python library for character-encoding detection, was released under the MIT licence after twenty years under the GNU Lesser General Public License (LGPL).
Its maintainer presented the release as a ground-up rewrite produced with an LLM coding agent, and the original author objected that a rewrite by people exposed to the LGPL code remains bound by that licence~\cite{chardet327}.
In response, the maintainer reported that JPlag found a maximum similarity of 1.29\% between the new release and the previous one~\cite{willison2026chardet,lwn2026chardet}.
Without taking a position on that dispute, we observe that token-similarity scores have become the evidence offered when an agent-assisted rewrite is claimed to be independent of the original, which leaves maintainers, downstream users and licence auditors to decide what such a score is worth.

Detectors such as JPlag and MOSS~\cite{prechelt2002jplag,schleimer2003} were built to catch copying that survives renaming and reordering, and a low score has long been read as support for independent authorship.
That reading has become harder to defend since LLM paraphrase was shown to lower token overlap while behaviour and structure persist~\cite{saglam2025aiplagiarism,zhang2024clonellm,park2025paraphrase,meeus2026functional}, and since coding agents extended rewriting to whole repositories, where they plan, edit files, run tests and repair failures with little supervision~\cite{jimenez2024swebench,yang2024sweagent,dong2025surveyagents}.
Existing studies cover parts of this situation without measuring it as a whole.
Paraphrase studies evaluate detectors on methods and student assignments~\cite{saglam2025aiplagiarism,zhang2024clonellm,park2025paraphrase}, memorisation and copyleft studies measure the leakage of protected fragments in short completions~\cite{yu2023codeipprompt,colombo2025copyleft,alkaswan2024traces}, and repository-rebuild benchmarks change the language or withhold the source and score behaviour only~\cite{repozero2026,nl2repo2025}.
None reports how similar a same-language, repository-scale rewrite is to its original, whether that similarity is reproducible from one run to the next, or how it compares with an independent implementation of the same task.

We measure these three quantities in the setting least favourable to a claim of independence, which we call \emph{source-conditioned rewriting}, where the agent works with the original in its workspace.
This choice bounds what a score can show, since a low score cannot rule out access to the original if rewrites produced with full access also score low.
Under our protocol, an operator gives the agent a working copy of the project, pastes a rewriting prompt followed by a repair prompt, and then sends only \texttt{Continue} until a predefined behavioural harness passes, without ever editing the code.
We apply this protocol to six copyleft Python projects with five agents from different providers, which yields 69 rewrites.
Fifteen of them cross the five agents with three prompts of increasing structural change on one project, and the remaining 54 repeat the most disruptive prompt, which requests a ground-up rearchitecture, three times with three agents on each of the six projects.

This paper makes the following contributions.
\begin{itemize}[leftmargin=*,itemsep=0.15em]
  \item We define an agent-only rewriting protocol that accounts separately for rewriting and for repair, and we release the 69 rewrites, their harnesses and all scores.
  \item We show that rewriting is cheap, since all 15 controlled rewrites pass the harness after agent-only repair and the initial rewrite takes a median of 18 minutes. Median similarity is 64\% under a layout-preserving prompt and 22\% under the ground-up prompt, although this ordering holds for three of the five agents only.
  \item We show that low similarity is common but not reproducible. Of the 54 ground-up rewrites, 42 score below 25\% mean JPlag similarity, 34 of these also pass the harness, and six score below 5\%, yet the nine rewrites of a single project spread over 14 to 75 percentage points. Repeating an identical configuration accounts for 61\% of the variance, against 14\% for the project and under 1\% for the agent, and in 7 of the 18 project-agent configurations three identical runs fall on both sides of the 25\% threshold.
  \item We show that two signals which resemble provenance are not specific to the original. Rewrites retain 54\% of the original's external imports and 48\% of its fingerprint facts, which is 44 and 35 percentage points more than unrelated projects do. When rewrites are compared with independent implementations that we wrote for each project from a capability specification and the harness, their fingerprint advantage is positive on only two of six projects, and neither signal separates the two groups (AUC 0.45 for imports and 0.40 for fingerprints, with confidence intervals that include 0.5).
\end{itemize}

For practice, these results mean that a similarity score from one rewrite is a single draw from a wide distribution, so that a low value neither demonstrates independence nor guarantees that the next run will score low.
They also mean that a candidate provenance signal must be compared with independent implementations of the same task, because a comparison with unrelated projects credits to the original what the task itself imposes.

\section{Background}
\label{sec:background}

The question we study arises where three topics meet.
Copyleft licences determine what a rewrite must achieve to be distributed under other terms, token-based detectors supply the evidence invoked to show that it does, and coding agents have made such rewrites cheap enough for the matter to arise in practice.
We present each in turn and then fix the vocabulary used in the rest of the paper.

\subsection{Copyleft and Independent Reimplementation}

A copyleft licence grants the right to use, modify and redistribute a program on the condition that derived works are distributed under the same terms.
The GNU GPL~\cite{gplv3} applies this condition to distributed copies, and the GNU AGPL~\cite{agplv3} extends it to software offered over a network.
Five of the six projects we study are under AGPL-3.0 and one is under GPL-3.0.
Licences do change during the life of a project~\cite{dipenta2010licensing}, and a study of 16,221 Java projects on GitHub found that the reasons for such changes are seldom recorded~\cite{vendome2017license}.
Moving copyleft code to a permissive licence is nevertheless commonly understood to require either the agreement of its copyright holders or new code that is not derived from the old.

The established way to produce such code is clean-room engineering, in which one team studies the reference system and writes a specification while a second team, which never reads the reference source, implements from that specification.
In {Computer Associates v.\ Altai}~\cite{altai1992}, a rewrite produced in this manner was held not to infringe, and the court compared the two programs only after filtering out the elements dictated by efficiency or by external constraints.
A coding agent removes the separation on which the practice rests, since the same loop that writes the new code can read the original whenever it is public, named in the prompt or present in the workspace.
Whether the result is a derived work is a legal question that this paper does not address.
We study the technical evidence that the parties to such a question put forward.

\subsection{Token-Based Similarity}

JPlag~\cite{prechelt2002jplag} converts each program into a stream of syntactic tokens, covers pairs of streams with common token runs above a minimum length, and reports the share of tokens covered.
MOSS~\cite{schleimer2003} selects fingerprints from hashed $k$-grams and compares the fingerprint sets.
Both abstract from layout, comments and identifier names, so that their scores survive renaming and reformatting, and recent work extends this resilience to inserted and reordered statements~\cite{saglam2024tokennorm}.

In the clone-detection taxonomy, these tools address textual clones, that is, exact copies (Type~1), copies with renamed identifiers (Type~2) and copies with changed, added or removed statements (Type~3)~\cite{bellon2007clone}.
Code that implements the same functionality with a different syntactic structure, referred to as Type~4 in later surveys~\cite{roy2009comparison}, lies outside their design.
A ground-up rewrite that keeps behaviour is an attempt to turn a program into a Type-4 clone of itself, so a low token score after such a rewrite shows that the surface changed and is silent on how the new code was obtained.

No agreed threshold separates copied from independent code.
A study of method-level completions treats a JPlag score of 0.70 as a match~\cite{colombo2025copyleft}, whereas the \texttt{chardet} maintainer cited a score close to 1\% as evidence of independence.
We report scores over their full range and state our operational threshold in Section~\ref{sec:setup}.

\subsection{Coding Agents}

A coding agent couples a language model with tools for reading and editing files and for running commands, and it iterates on the feedback it observes~\cite{dong2025surveyagents}.
SWE-bench made the repair of real repositories under their test suites the reference task for such systems~\cite{jimenez2024swebench}, and agents such as SWE-agent~\cite{yang2024sweagent} and OpenHands~\cite{wang2024openhands} now carry out multi-file edits and repairs without supervision.
Our protocol relies on this loop as shipped in commodity products and adds nothing to it.

\subsection{Terminology}

The \emph{original} is the copyleft project as published.
A \emph{rewrite} is a repository produced by an agent that was asked to reimplement the original in the same language.
Rewriting is \emph{source-conditioned} when the original is present in the agent's workspace, which is the only setting we study.
An \emph{independent implementation} is a program that fulfils the same task and was written without the original at hand, and it serves as our control.
The \emph{harness} is the fixed set of executable checks that decides whether a rewrite or a control behaves like the original.
A \emph{run} is one execution of the protocol for a given project, agent and prompt.

\section{Study Design}
\label{sec:setup}

We study source-conditioned rewriting, the setting least favourable to a claim of independence, in which the agent has the original in its workspace.
If rewrites produced under full access still score low, a low score cannot rule out access. We ask three questions about such rewrites: whether they are cheap to obtain, whether their similarity score is reproducible, and whether provenance signals can tell them from independent code. 
This section defines the questions, the corpus, the protocol, the measures and the controls, in that order.

\subsection{Research Questions}

\begin{description}[leftmargin=1.5em,itemsep=0.25em]
  \item[RQ1 (feasibility).] \textit{Can a coding agent, with no human edit to the code, rewrite a copyleft repository so that it passes a fixed behavioural harness, and at what cost?} To answer this research question, we run five agents under three prompts of increasing structural change. For each run we record the harness outcome before and after repair, the session time, the dialogue tokens, the size of the rewrite and its JPlag~\cite{prechelt2002jplag} similarity to the original. This establishes that rewriting is practical and shows how the prompt shapes similarity. The three prompts differ in wording, paradigm and layout freedom at once, so we attribute effects to the prompt as a whole.
  \item[RQ2 (stability).] \textit{How much does similarity to the original vary across repeated rewrites under a fixed prompt, and how often does this variation change a threshold verdict?} In this research question, we repeat the ground-up prompt from RQ1 three times with three agents on each of the considered projects. We split the variance of the similarity score into project, agent, their interaction and repetition. We also count the configurations whose identical runs fall on both sides of a threshold. This tells how much a single published score can be trusted.
  \item[RQ3 (calibration).] \textit{Do import and fingerprint retention distinguish a rewrite from an independent implementation of the same task?} For this last research question, we measure how many of the original's external imports and fingerprint facts each rewrite retains. We compare this retention with two references: the other campaign originals, and a same-task independent control per project. We then test whether either signal separates the rewrites from the controls. A signal that only beats unrelated projects may be measuring the task and not the origin.
\end{description}

\subsection{Overview}

Table~\ref{tab:design} summarises the design.
The controlled part (RQ1) crosses five agents with three prompts on one project.
The replication part (RQ2, RQ3) repeats the ground-up prompt three times with three agents on six projects.
The two parts share no run, for 69 rewrites in total. RQ2 and RQ3 use the ground-up prompt because it gave the lowest median similarity in RQ1, which is the case most favourable to a claim of independence.
This choice rests on one run per cell in RQ1, a limit we discuss in Section \ref{sec:discussion}.

\begin{table}[!h]
\caption{Study design. A run is one execution of the protocol for a project, an agent and a prompt.}
\label{tab:design}
\small
\begin{tabular}{@{}lrrrrrl@{}}
\toprule
\textbf{Part} & \textbf{Projects} & \textbf{Agents} & \textbf{Prompts} & \textbf{Repetitions} & \textbf{Runs} & \textbf{Used for} \\
\midrule
Controlled  & 1 & 5 & 3 & 1 & 15 & RQ1 \\
Replication & 6 & 3 & 1 & 3 & 54 & RQ2, RQ3 \\
\bottomrule
\end{tabular}
\end{table}

\subsection{Projects}

Candidates are copyleft Python repositories collected through the GitHub GraphQL API, capped at 15\,000 Python SLOC and ranked by GitHub star count (descending), from a snapshot taken on 2 June 2026.
Documentation-only repositories are excluded.
We retained the six projects of Table~\ref{tab:subjects} as the first six entries of that Python ranking.
The selection is purposive, not random.
\emph{Deep-Live-Cam}, the top-ranked candidate, is the project of the controlled part.
Five projects are under AGPL-3.0 and one under GPL-3.0.

\begin{table}[h]
\caption{Subject systems. SLOC counts Python lines excluding blanks and comment-only lines.}
\label{tab:subjects}
\small
\begin{tabular}{@{}llrrp{0.36\linewidth}@{}}
\toprule
\textbf{Name} & \textbf{Licence} & \textbf{Files} & \textbf{SLOC} & \textbf{Description} \\
\midrule
\emph{Deep-Live-Cam} & AGPL-3.0 & 33 & 5787 & Real-time face swap and one-click video deepfake from a single image \\
\emph{ChatTTS} & AGPL-3.0 & 65 & 9951 & Generative speech model for daily dialogue \\
\emph{PDFMathTranslate} & AGPL-3.0 & 32 & 6396 & PDF scientific-paper translation with preserved layout \\
\emph{algo} & AGPL-3.0 & 6 & 911 & Personal VPN deployment in the cloud \\
\emph{agenticSeek} & GPL-3.0 & 44 & 6522 & Fully local autonomous agent for browsing and coding \\
\emph{heretic} & AGPL-3.0 & 29 & 5084 & Automatic censorship removal for language models \\
\bottomrule
\end{tabular}
\end{table}

\subsection{Agents}

An agent is a model driven through GitHub Copilot agent mode in Visual Studio Code with default settings.
The controlled part uses Claude Haiku~4.5, DeepSeek~V4~Pro, Gemini~2.5~Pro, GPT-5.4~mini and Qwen~3.7~Plus.
The replication part keeps Claude, Gemini and Qwen, chosen after the controlled part to cover three providers at manageable cost. 
Runs took place between 17 August and 30 September 2026.
Model identifiers are given in the replication package.

\subsection{Rewriting Protocol}
\label{sec:protocol}

The operator runs every session by pasting fixed prompts and never edits code or comments on a failure. No prompt asks the agent to evade a licence or hide a copy.
Each run starts from a fresh working copy of the original on disk, opened as the Visual Studio Code workspace. The operator selects the target model in GitHub Copilot and pastes the rewriting prompt into a new agent-mode chat on that workspace. The agent edits the local tree in place. Repair uses a second new chat on the same working copy. 

\begin{enumerate}[leftmargin=*,itemsep=0.2em]
  \item \emph{Rewriting.}
  The operator opens a working copy of the original and pastes one of three prompts.
  S1 (layout-preserving) requests a line-level rewrite with new identifiers and the same file layout.
  S2 (paradigm shift) additionally requires a change of programming paradigm within the same modules.
  S3 (ground-up) requests a rearchitecture with free file organisation.
  The agent edits the working copy in place.
  If it stops before completion, the operator sends \texttt{Continue}.
  \item \emph{Cleanup (S3 only).}
  A fixed prompt asks the agent to remove the files that the new layout supersedes.
  \item \emph{Testing and repair.}
  The operator pastes the project kit (harness, ordered test cases, repair prompt) in a fresh dialogue.
  The agent runs the harness and patches the rewrite until all cases pass.
  Changes are restricted to integration code (imports, aliases, entry points, paths, dependency wiring).
  The agent may read the original but may edit only the rewrite, and it may neither copy modules of the original nor call it at run time.
  A final prompt removes auxiliary files created during repair.
  \item \emph{Archiving.}
  We archive the tree before and after step~3, and we check that no file of the original remains unchanged and that no entry point imports or invokes the original.
\end{enumerate}

Because the agent edits in place, an archived rewrite may contain files of the original that were lightly modified or not removed.
We score the tree as delivered and examine leftover files in the appendix. 

\subsection{Behavioural Harness}

The harness of a project is a fixed suite of executable cases, six for \emph{Deep-Live-Cam} and five for the others.
Each case runs the original and a candidate on the same input and compares the outputs with a kind-specific distance (image, video, text, PDF or binary) and fixed margins. The case fails if the candidate is too close to the untransformed input or too far from the original's output relative to the gap between that input and that output.
A candidate passes when all cases succeed.
A pass shows agreement on these cases only, since the harness executes 20\% to 58\% of the lines of the originals (appendix).

\subsection{Similarity}
\label{sec:similarity}

We run JPlag~\cite{prechelt2002jplag} version 6.3.0 with the default Python minimum token match of 12 on the Python sources of an original $O$ and a rewrite $C$, excluding tests, virtual environments, installed packages, caches, version-control metadata, the repair kit and root-level asset directories.
$\mathrm{sim}(O,C)$ is the mean of the two directional similarities reported by JPlag, and $\mathrm{sim}_{\max}(O,C)$ is the larger.
Scores are taken after testing and repair unless stated otherwise.
The appendix reports five other measures on all 69 rewrites (copydetect, Dolos, MOSS, PMD CPD, line-level Jaccard).

The joint criterion at threshold $\tau$ is
\[
J_\tau(C) \;=\; \bigl[\mathrm{sim}(O,C) < \tau\bigr] \;\wedge\; \mathrm{pass}(C).
\]
No threshold is established for repository-level comparison (Section~\ref{sec:background}).
We use $\tau = 25\%$\footnote{Unrelated rewrites in our corpus score at most 10.5\% against each other (Section~\ref{sec:rq3}), so the three thresholds lie above the level of unrelated code.} and repeat threshold-dependent results at 15\% and 30\%.

\subsection{Retention Signals}
\label{sec:retention}

We extract from each Python file in scope the roots of external imports, called API names, keyword-argument constants, distinctive string and numeric literals, and named configuration values.
Standard-library and first-party imports are excluded.
$\mathrm{Obs}(X)$ denotes the facts of repository $X$.
For a set $A$ of facts of the original, retention in $X$ is
\[
\mathrm{Ret}(A,X) \;=\; \frac{|A \cap \mathrm{Obs}(X)|}{|A|}.
\]
$A$ is either $\mathrm{Imp}(O)$, the external import roots of $O$, or the fingerprint $F(O)$, the facts of $O$ selected by a fixed heuristic that keeps all external imports, up to 40 distinctive strings, 40 API names, 20 numeric literals, and all named configuration bindings (97 to 172 facts per project).
The fingerprint has two layers, $F_{\mathrm{atom}}$ for individual facts and $F_{\mathrm{cg}}$ for call-graph edges from local functions to APIs.
$F(O)$ is that heuristic set. Separately, $F_{\mathrm{atom}}$ is a larger IDF-selected atomic layer (capped at 400 facts, dropping facts that appear in every original), used with $F_{\mathrm{cg}}$ for the STIC contrasts. The two sets are not identical.

The margin compares a rewrite with the five other originals.
\[
M(A,C) \;=\; \mathrm{Ret}(A,C) \;-\; \frac{1}{5}\sum_{O' \neq O} \mathrm{Ret}(A,O')
\]
The appendix reports execution-stage coverage and relational and structural fingerprints.

\subsection{Independent Controls}
\label{sec:controls}

The margin $M$ credits a rewrite with facts that the task imposes, such as the libraries that any face-swapping tool imports.
For each project we therefore built one \emph{same-task independent control} (STIC), denoted $S_O$.
One author implemented the project from a capability specification and the harness, without copying from the original, and with a different package layout, command-line interface and identifiers.
A control is accepted when it passes the harness and scores 0\% mean and maximum JPlag similarity against the original.
Third-party libraries required by the task are allowed.
The control for \emph{algo} omits the Ansible deployment of remote servers, which the harness does not exercise.

We compare each rewrite with its control.
\[
\Delta_{\mathrm{STIC}}(A,C) \;=\; \mathrm{Ret}(A,C) \;-\; \mathrm{Ret}(A,S_O)
\]
A positive value means that the rewrite retains more of the original than an independent implementation of the same task.
We also report the precision $|A \cap \mathrm{Obs}(X)|/|\mathrm{Obs}(X)|$ to expose size effects.

The controls have three limits.
There is one per project, written by one person.
They were extended after the campaign.
Their first versions passed the harness with much less code than the rewrites, so the same author added modules to all six under the same acceptance conditions.
This took place after the first retention results had been examined, although the author consulted neither $F(O)$ nor the per-rewrite contrasts.
We report retention before and after extension in Section~\ref{sec:rq3}.
Lastly, the 0\% acceptance condition prevents their use to calibrate a JPlag threshold.

\subsection{Analysis}
\label{sec:analysis}

RQ1 reports, per run, the harness outcome and the similarity before and after repair, the session time and the dialogue tokens (prompt plus completion sizes in the session log).
RQ2 is descriptive, given three repetitions per configuration.
We report the median and range per project, the shares of the total sum of squares due to project, agent, their interaction and repetition, and the number of configurations whose three runs fall on both sides of $\tau$.
Proportions carry Wilson 95\% intervals, and the effect of repair is tested on paired scores with a Wilcoxon signed-rank test.
RQ3 reports $M$ and $\Delta_{\mathrm{STIC}}$ per project, and the area under the ROC curve for separating the 54 rewrites from the six controls, with 95\% intervals from 2,000 bootstrap resamples.

\section{Results}
\label{sec:results}

We answer the three questions in turn.
Similarity is $\mathrm{sim}(O,C)$ after testing and repair unless stated otherwise.

\subsection{RQ1: Feasibility and Cost}
\label{sec:rq1}

\paragraph{Harness outcome.}
All 15 rewrites of \emph{Deep-Live-Cam} pass the harness at the end of the protocol (Table~\ref{tab:rq1-exec}, Wilson 95\% interval 80\% to 100\%).
Fourteen fail it after the rewriting step and are repaired by the agent, and one (DeepSeek, S2) passes without repair.
The operator sent a median of three \texttt{Continue} messages per rewriting session and made no other intervention.

\paragraph{Cost.}
The rewriting session takes a median of 18 minutes (range 8 to 125), and 14 of 15 sessions end within an hour.
The median rises with the prompt, from 9 minutes under S1 to 30 under S2 and 38 under S3.
The repair session takes a median of 15 minutes (range 5 to 59).
Rewriting uses 13.1 million dialogue tokens over the 15 runs (median 0.52 million per run), and repair uses 20.0 million (median 0.95 million).
Repair uses more tokens than rewriting in 10 of 15 runs.
The rewrites contain 1{,}673 to 9{,}648 SLOC, against 5{,}787 for the original.

\begin{figure}[t]
\centering
\includegraphics[width=\linewidth]{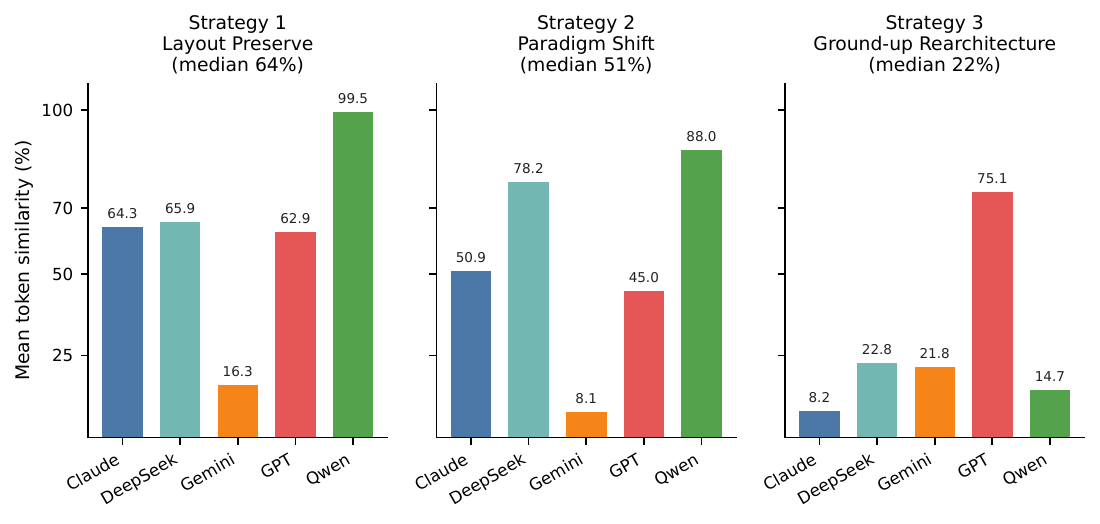}
\caption{RQ1. Similarity to \emph{Deep-Live-Cam} after repair, by prompt and agent. Each bar is one run. The dashed line marks $\tau = 25\%$.}
\label{fig:rq1-jplag}
\end{figure}

\paragraph{Similarity by prompt.}
Median similarity is 64\% under S1, 51\% under S2 and 22\% under S3 (Figure~\ref{fig:rq1-jplag}).
The decrease is not uniform across agents.
For Claude, DeepSeek and Qwen, S3 yields the lowest score, at least 43 points below S1.
Gemini scores between 8\% and 22\% under all three prompts.
GPT scores 75\% under S3, its highest value, which we attribute to files of the original that remain in the archived tree next to the new layout, with imports between the two.
With one run per cell, these differences cannot be separated from run-to-run variation, which RQ2 quantifies.

Six rewrites score below $\tau = 25\%$, four under S3 and two from Gemini under S1 and S2.
Since all rewrites pass the harness, the same six meet $J_{25}$.
Three meet $J_{15}$ and six meet $J_{30}$.

\paragraph{Effect of repair.}
Repair changes similarity by a median of 0.4 points in absolute value (range $-0.2$ to $+4.1$), and no rewrite crosses $\tau$ during repair.

\begin{table}[h]
\caption{RQ1. The 15 rewrites of \emph{Deep-Live-Cam}. All pass the harness after repair.
SLOC excludes blank and comment-only lines.
Rewriting and repair minutes sum the agent run times of each session and exclude idle time between operator messages.
Similarity is mean/max JPlag before and after repair.
Tokens are dialogue tokens of the rewriting and repair sessions.}
\label{tab:rq1-exec}
\setlength{\tabcolsep}{2.5pt}
\resizebox{\linewidth}{!}{%
\begin{tabular}{@{}llrrrrrrr@{}}
\toprule
& & & \multicolumn{2}{c}{\textbf{Minutes}} & \multicolumn{2}{c}{\textbf{Similarity}} & \multicolumn{2}{c}{\textbf{Tokens}} \\
\cmidrule(lr){4-5}\cmidrule(lr){6-7}\cmidrule(lr){8-9}
\textbf{Agent} & \textbf{Prompt} & \textbf{SLOC} & \textbf{Rewriting} & \textbf{Repair} & \textbf{Before} & \textbf{After} & \textbf{Rewriting} & \textbf{Repair} \\
\midrule
Claude Haiku 4.5 & S1 & 3{,}604 & 9 & 22 & 64.5/87.0 & 64.3/86.1 & 116\,386 & 1\,428\,312 \\
Claude Haiku 4.5 & S2 & 5{,}364 & 31 & 15 & 49.9/51.0 & 50.9/53.4 & 794\,347 & 946\,136 \\
Claude Haiku 4.5 & S3 & 5{,}472 & 18 & 6 & 8.4/11.0 & 8.2/11.1 & 316\,776 & 419\,516 \\
DeepSeek V4 Pro & S1 & 4{,}051 & 37 & 31 & 65.7/83.5 & 65.9/84.6 & 1\,379\,263 & 941\,589 \\
DeepSeek V4 Pro & S2 & 9{,}221 & 30 & 7 & 77.2/99.5 & 78.2/99.5 & 157\,625 & 316\,931 \\
DeepSeek V4 Pro & S3 & 4{,}855 & 38 & 33 & 22.7/28.2 & 22.8/27.7 & 360\,725 & 2\,728\,838 \\
Gemini 2.5 Pro & S1 & 2{,}912 & 8 & 49 & 15.9/28.4 & 16.3/29.7 & 378\,741 & 4\,515\,401 \\
Gemini 2.5 Pro & S2 & 1{,}946 & 13 & 23 & 6.7/15.5 & 8.1/17.3 & 2\,033\,816 & 1\,720\,140 \\
Gemini 2.5 Pro & S3 & 1{,}673 & 17 & 59 & 20.2/44.5 & 21.8/45.2 & 220\,574 & 2\,185\,001 \\
GPT-5.4 mini & S1 & 3{,}341 & 13 & 8 & 63.0/78.7 & 62.9/78.8 & 336\,330 & 347\,559 \\
GPT-5.4 mini & S2 & 3{,}512 & 11 & 7 & 45.2/55.2 & 45.0/56.1 & 519\,638 & 342\,001 \\
GPT-5.4 mini & S3 & 7{,}953 & 47 & 23 & 71.0/89.1 & 75.1/89.0 & 2\,494\,556 & 717\,790 \\
Qwen 3.7 Plus & S1 & 5{,}813 & 9 & 14 & 99.0/99.6 & 99.5/99.5 & 756\,613 & 915\,853 \\
Qwen 3.7 Plus & S2 & 8{,}126 & 125 & 5 & 87.4/93.5 & 88.0/93.0 & 2\,710\,061 & 1\,138\,933 \\
Qwen 3.7 Plus & S3 & 9{,}648 & 52 & 13 & 14.6/14.7 & 14.7/14.7 & 552\,785 & 1\,293\,895 \\
\midrule
Total & & 77{,}491 & 458 & 316 & -- & -- & 13.1M & 20.0M \\
\bottomrule
\end{tabular}%
}
\end{table}

\find{{\bf [RQ-1] \ding{42}}
With no human edit to the code, a coding agent rewrote \emph{Deep-Live-Cam} to a harness pass in all 15 runs, after rewriting and repair sessions of 18 and 15 minutes at the median, respectively.
Six of the 15 rewrites also score below 25\% similarity, four of them under the ground-up prompt.
This evidence comes from one project and one run per configuration.
}

\subsection{RQ2: Stability of Similarity}
\label{sec:rq2}

\paragraph{Harness outcome and cost.}
Of the 54 rewrites, 46 pass the harness (Wilson 95\% interval 73\% to 92\%).
Six of the eight failures are on \emph{ChatTTS}, whose harness compares generated speech with that of the original (three Claude runs at 1/5 cases, three Qwen runs at 0/5).
The other two are Claude runs on \emph{PDFMathTranslate} (4/5) and \emph{Heretic} (3/5).
Gemini passes in all 18 runs.
The rewriting session takes a median of 30 minutes (5.5 to 160) and the repair session a median of 20 minutes (1 to 191).
Over the 54 runs, rewriting uses 338 million dialogue tokens and 33.1 hours, and repair uses 217 million tokens and 39.4 hours.
Per-run figures are in the appendix.

\paragraph{Similarity level.}
Median similarity over the 54 rewrites is 12.6\% (interquartile range 6.5\% to 24.0\%).

Forty-two rewrites score below $\tau = 25\%$ (78\%, interval 65\% to 87\%), and 34 of them also pass the harness and so meet $J_{25}$ (63\%, interval 50\% to 75\%).
The counts are 29 and 26 at $\tau = 15\%$, and 47 and 39 at $\tau = 30\%$.
Six rewrites score below 5\%.

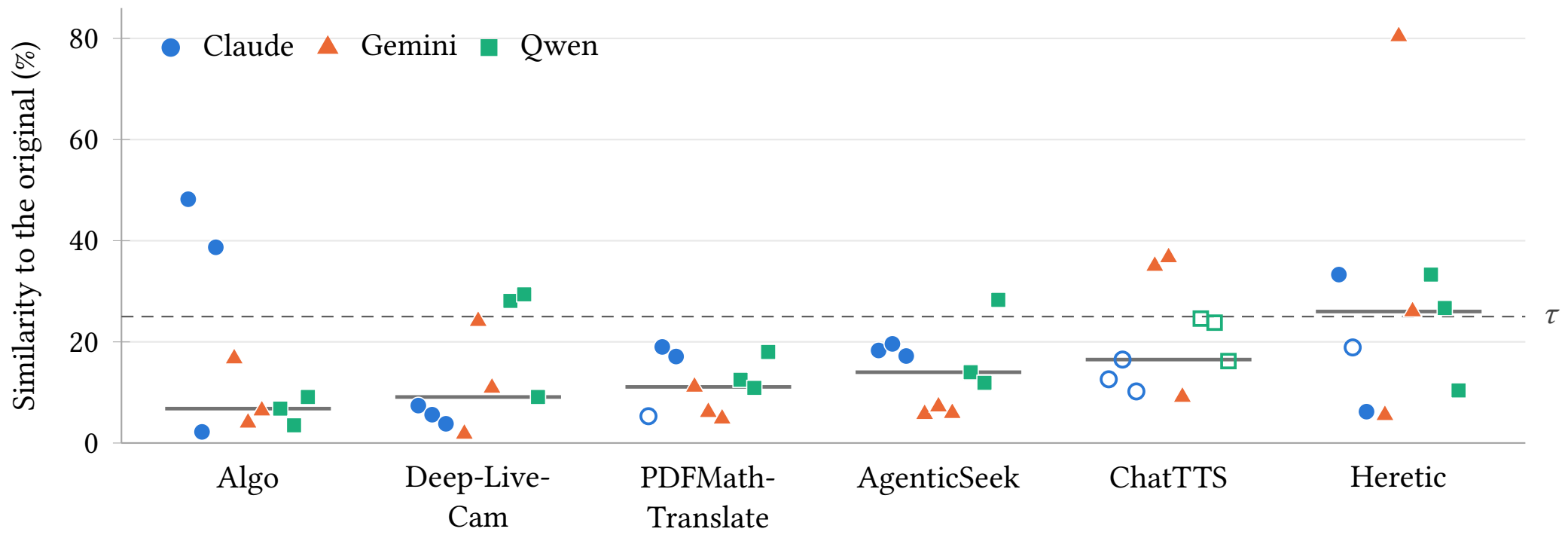
\begin{figure}[t]
\centering
\definecolor{agA}{HTML}{2A78D6}
\definecolor{agB}{HTML}{EB6834}
\definecolor{agC}{HTML}{1BAF7A}
\begin{tikzpicture}
\begin{axis}[
  width=\linewidth, height=5.4cm,
  xmin=0.45, xmax=6.55, ymin=0, ymax=86,
  xtick={1,...,6},
  xticklabels={Algo, Deep-Live-\\Cam, PDFMath-\\Translate, AgenticSeek, ChatTTS, Heretic},
  xticklabel style={font=\footnotesize, align=center},
  ytick={0,20,40,60,80},
  yticklabel style={font=\footnotesize},
  ylabel={Similarity to the original (\%)},
  ylabel style={font=\footnotesize},
  axis lines=left, axis line style={-, gray!70},
  tick style={gray!70}, xtick style={draw=none},
  ymajorgrids, grid style={gray!18},
  clip=false,
  legend style={at={(0.02,0.98)}, anchor=north west, draw=none, fill=none,
                font=\footnotesize, legend columns=3, column sep=4pt},
  legend image post style={mark options={solid}},
]
\draw[densely dashed, black!65] (axis cs:0.45,25) -- (axis cs:6.55,25);
\node[anchor=west, font=\footnotesize, black!75] at (axis cs:6.57,25) {$\tau$};
\foreach \x/\m in {1/6.8, 2/9.1, 3/11.1, 4/14.0, 5/16.5, 6/26.0}{
  \edef\tmp{\noexpand\draw[black!55, line width=0.9pt]
    (axis cs:\x-0.36,\m) -- (axis cs:\x+0.36,\m);}\tmp
}
\addplot[only marks, mark=*, mark size=2.1pt, agA, mark options={draw=white, line width=0.3pt}]
  coordinates {(0.74,48.2) (0.80,2.2) (0.86,38.7)
               (1.74,7.4) (1.80,5.6) (1.86,3.8)
               (2.80,19.0) (2.86,17.1)
               (3.74,18.3) (3.80,19.6) (3.86,17.2)
               (5.74,33.3) (5.86,6.2)};
\addlegendentry{Claude}
\addplot[only marks, mark=triangle*, mark size=2.7pt, agB, mark options={draw=white, line width=0.3pt}]
  coordinates {(0.94,16.7) (1.00,4.0) (1.06,6.4)
               (1.94,1.8) (2.00,24.1) (2.06,10.9)
               (2.94,11.1) (3.00,6.1) (3.06,4.8)
               (3.94,5.7) (4.00,7.2) (4.06,5.9)
               (4.94,35.0) (5.00,36.7) (5.06,9.1)
               (5.94,5.5) (6.00,80.4) (6.06,26.0)};
\addlegendentry{Gemini}
\addplot[only marks, mark=square*, mark size=1.9pt, agC, mark options={draw=white, line width=0.3pt}]
  coordinates {(1.14,6.8) (1.20,3.5) (1.26,9.1)
               (2.14,28.1) (2.20,29.4) (2.26,9.1)
               (3.14,12.5) (3.20,10.9) (3.26,18.0)
               (4.14,14.0) (4.20,11.9) (4.26,28.3)
               (6.14,33.3) (6.20,26.7) (6.26,10.4)};
\addlegendentry{Qwen}
\addplot[only marks, mark=o, mark size=1.9pt, agA, line width=0.7pt, forget plot]
  coordinates {(2.74,5.3) (4.74,12.6) (4.80,16.5) (4.86,10.2) (5.80,18.9)};
\addplot[only marks, mark=square, mark size=1.7pt, agC, line width=0.7pt, forget plot]
  coordinates {(5.14,24.6) (5.20,23.8) (5.26,16.2)};
\end{axis}
\end{tikzpicture}
\caption{RQ2. Similarity after repair of the 54 ground-up rewrites, by project and agent.
Each marker is one run, and hollow markers are runs that fail the harness.
Horizontal bars are project medians, and the dashed line marks $\tau = 25\%$.
Projects are ordered by median.}
\label{fig:rq2-strip}
\end{figure}

\paragraph{Stability across runs.}
Scores for a single project spread over 14 points on \emph{PDFMathTranslate} and up to 75 points on \emph{Heretic} (Figure~\ref{fig:rq2-strip}, Table~\ref{tab:rq2-summary}).
On \emph{Algo}, the three Claude runs alone give 48.2\%, 2.2\% and 38.7\%.
Repeating an identical configuration accounts for 61\% of the total sum of squares, against 14\% for the project, 25\% for the interaction between project and agent, and 0.1\% for the agent.
Neither the project nor the agent has a detectable effect (Kruskal-Wallis, $p = 0.23$ and $p = 0.33$).
The standard deviation between runs of one configuration is 13.2 points, which exceeds the median score.

This variation changes threshold verdicts.
In 7 of the 18 configurations, the three runs fall on both sides of $\tau = 25\%$, and five of the six projects have rewrites on both sides.
The count is 12 of 18 at $\tau = 15\%$ and 5 of 18 at $\tau = 30\%$.

\paragraph{Effect of repair.}
Repair has no systematic direction (median change 0.0 points, Wilcoxon signed-rank $p = 0.74$), but it moves individual scores.
Eighteen rewrites change by 10 points or more, and 18 cross $\tau = 25\%$ during repair, 11 downwards and 7 upwards.
On \emph{Algo}, three rewrites drop by 33 to 66 points because the agent restructured the code again during repair, beyond the integration changes that the prompt allows.
One \emph{Heretic} run rises from 11.2\% to 80.4\% 
On \emph{Deep-Live-Cam}, as in RQ1, no score moves by more than 2 points.

\begin{table}[t]
\caption{RQ2. Similarity (\%) of the 54 ground-up rewrites by project. A configuration is a project and an agent, and it straddles $\tau$ when its three runs fall on both sides of 25\%.}
\label{tab:rq2-summary}
\small
\begin{tabular}{@{}lrrrccc@{}}
\toprule
\textbf{Project} & \textbf{Median} & \textbf{Min} & \textbf{Max} & \textbf{Below 25\%} & \textbf{Straddling} & \textbf{Pass} \\
\midrule
\emph{Deep-Live-Cam}    &  9.1 & 1.8 & 29.4 & 7/9 & 1/3 & 9/9 \\
\emph{Algo}             &  6.8 & 2.2 & 48.2 & 7/9 & 1/3 & 9/9 \\
\emph{AgenticSeek}      & 14.0 & 5.7 & 28.3 & 8/9 & 1/3 & 9/9 \\
\emph{ChatTTS}          & 16.5 & 9.1 & 36.7 & 7/9 & 1/3 & 3/9 \\
\emph{PDFMathTranslate} & 11.1 & 4.8 & 19.0 & 9/9 & 0/3 & 8/9 \\
\emph{Heretic}          & 26.0 & 5.5 & 80.4 & 4/9 & 3/3 & 8/9 \\
\midrule
All                     & 12.6 & 1.8 & 80.4 & 42/54 & 7/18 & 46/54 \\
\bottomrule
\end{tabular}
\end{table}

\find{{\bf [RQ-2] \ding{42}}
Under the ground-up prompt, 42 of 54 rewrites score below 25\% similarity and 34 also pass the harness.
The score is not reproducible.
Repeating a configuration accounts for 61\% of the variance, and in 7 of 18 configurations identical runs fall on both sides of the threshold.
Repair has no systematic direction but shifts individual scores by up to 69 points.
}

\subsection{RQ3: Calibration of Retention Signals}
\label{sec:rq3}

\begin{figure}[h]
\centering
\includegraphics[width=\linewidth]{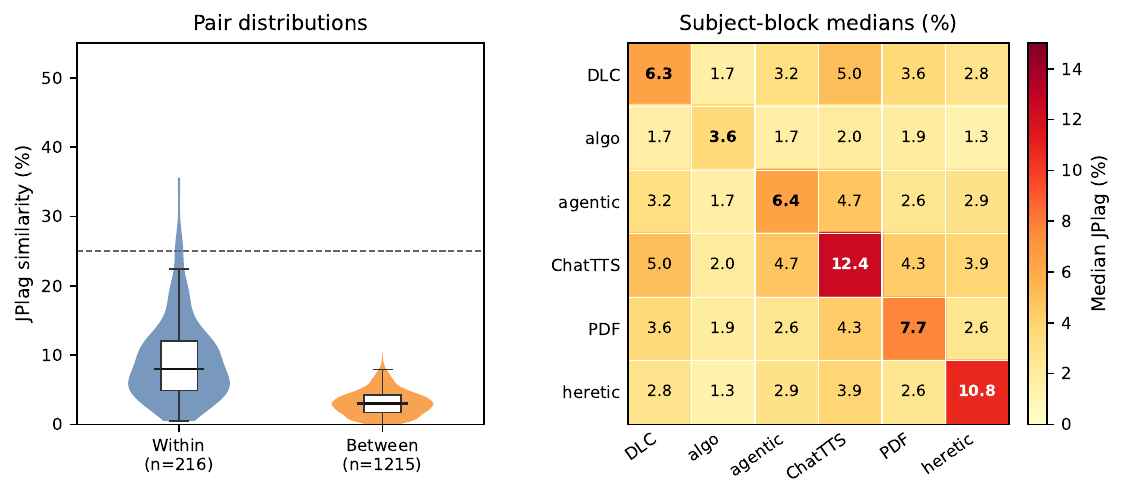}
\caption{RQ3. Similarity between rewrites. The left panel shows the distribution over the 216 within-project and 1{,}215 between-project pairs, with a dashed line at 25\%. The right panel shows the median similarity between rewrites of two projects, with same-project pairs on the diagonal.}
\Description{Left, violin plots of within-project and between-project pair similarities. Right, a six-by-six matrix of medians across projects.}
\label{fig:clone-between}
\end{figure}

\begin{figure}[h]
\centering
\includegraphics[width=\textwidth]{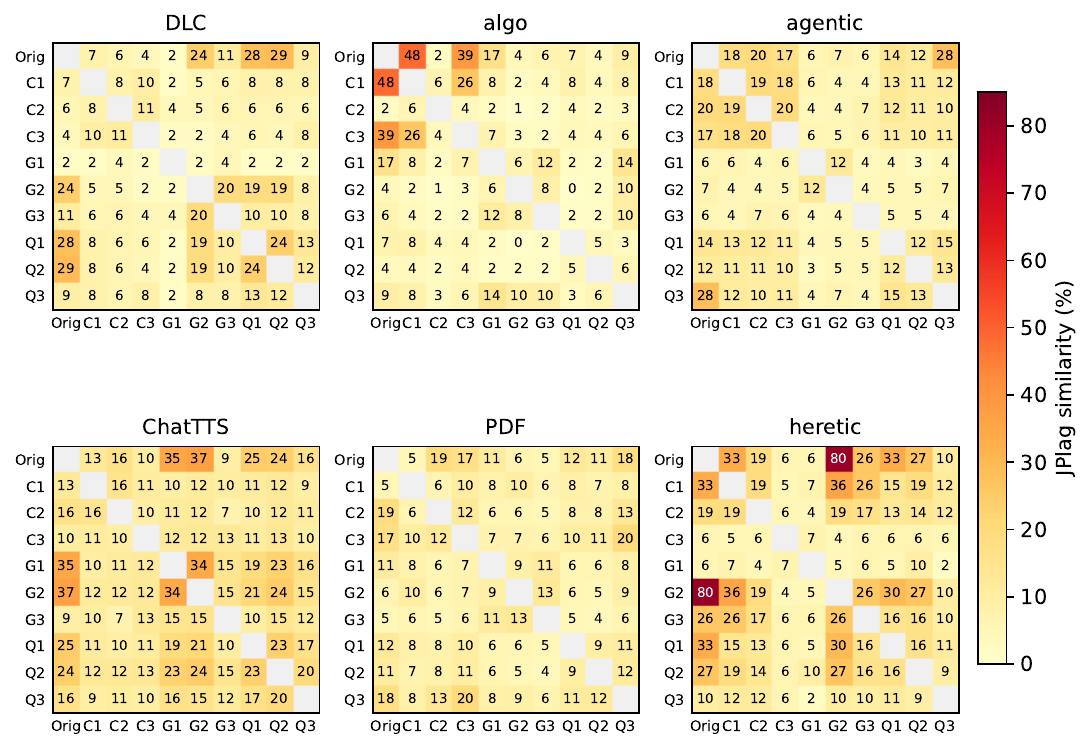}
\caption{RQ3. JPlag similarity (\%) among the original and the nine rewrites of each project. C, G and Q denote Claude, Gemini and Qwen, and the digit is the repetition.}
\Description{Six heatmaps, one per project. Each grid includes the original and its nine rewrites.}
\label{fig:clone-within}
\end{figure}

\paragraph{Similarity among rewrites.}
Two rewrites of different projects score a median of 3.0\% against each other and at most 10.5\% (1{,}215 pairs, Figure~\ref{fig:clone-between}).
This is the level that JPlag reports for unrelated code in this corpus.
Two rewrites of the same project score a median of 8.0\% (216 pairs, Figure~\ref{fig:clone-within}), and 7 pairs reach 25\%, 5 of them on \emph{Heretic}.
When both rewrites are below 15\% against their original (61 pairs), the median is 5.5\% and no pair reaches 25\%.
Rewrites of one original are thus almost as far from each other as from unrelated code, and token similarity does not reveal their common origin.
We therefore turn to imports and fingerprints.

\paragraph{Against unrelated projects.}
Rewrites retain on average 53.6\% of the external imports of their original and 47.7\% of its fingerprint (Table~\ref{tab:rq3-margins}).
The five unrelated originals retain 9.8\% and 12.7\% of the same sets, which gives margins of $+43.8$ and $+35.0$ points.
The margin is positive in every project.
It also holds when token similarity is low.
Among the 42 rewrites below $\tau = 25\%$, 81\% keep a fingerprint margin above 15 points, and 93\% retain more of the fingerprint than every unrelated original.
Read against this reference, both signals point to the original.

\paragraph{Against independent controls.}
The controls change this reading (Table~\ref{tab:rq3-stic-ret}).
They retain on average 49.7\% of the atomic fingerprint, against 44.5\% for rewrites, and 54\% of the imports, the same share as rewrites.
Mean $\Delta_{\mathrm{STIC}}$ on the atomic layer is positive in two projects, \emph{Algo} ($+31.5$) and \emph{ChatTTS} ($+6.2$), and negative in the other four, down to $-28.7$ on \emph{AgenticSeek}.
Over the 54 rewrites, 19 retain more of the atomic fingerprint than their control, and 24 more of the call-graph fingerprint.
Most of the facts that separate rewrites from unrelated projects are therefore facts that an implementation of the same task also contains.
Precision is of the same order for controls (0.10 to 0.24) and rewrites (0.17 to 0.29), so the controls do not owe their retention to size alone.

\paragraph{Retention follows token similarity.}
In all six projects, fingerprint retention rises with token similarity (appendix).
On \emph{Deep-Live-Cam}, the rank correlation between $\mathrm{sim}$ and $\Delta_{\mathrm{STIC}}$ over the nine rewrites is 0.97.
As a result, $\Delta_{\mathrm{STIC}}$ is positive for 3 of the 11 rewrites below 6\% similarity, 4 of the 18 between 6\% and 15\%, and 12 of the 25 above 15\%.
Without \emph{Algo}, these counts are 0 of 8, 1 of 15 and 9 of 22.
The fingerprint thus exceeds the control mostly where token similarity is already high.

\paragraph{Separation.}
Used as a score to separate the 54 rewrites from the six controls, atomic retention reaches an AUC of 0.40 (95\% interval 0.15 to 0.67) and import retention 0.45 (0.28 to 0.65).
Normalising within project gives 0.37 (0.14 to 0.64) and 0.45 (0.22 to 0.65).
All intervals include 0.5.
With six negatives the estimate is imprecise, so we report that we found no separation and do not claim that none exists.
JPlag separates the two groups by construction, since controls were required to score 0\%, and we do not count this as a result.

\paragraph{Sensitivity.}
Two choices affect these numbers.
The first is the extension of the controls.
It raised the atomic retention of the control from 33.2\% to 44.0\% on \emph{Deep-Live-Cam} and from 7.3\% to 17.8\% on \emph{Algo}.
With the first version, mean $\Delta_{\mathrm{STIC}}$ on \emph{Deep-Live-Cam} would be $+8.3$ points instead of $-2.5$, so the sign on this project depends on the extension.
The second is the scope of the \emph{Algo} control, which omits the Ansible layer and therefore cannot contain the facts of that layer.
Restricting the fingerprint to the one module that both cover (25 facts) gives a mean $\Delta_{\mathrm{STIC}}$ of $+25.8$ points, positive for 7 of 9 rewrites, so \emph{Algo} remains positive under both scopes.

\begin{table}[t]
\caption{RQ3. Mean retention (\%) of the original's imports and fingerprint in the nine rewrites of each project and in the five unrelated originals. $M$ is the difference in percentage points.}
\label{tab:rq3-margins}
\small
\begin{tabular}{@{}lrrrrrr@{}}
\toprule
 & \multicolumn{3}{c}{\textbf{Imports}} & \multicolumn{3}{c}{\textbf{Fingerprint}} \\
\cmidrule(lr){2-4}\cmidrule(l){5-7}
\textbf{Project} & \textbf{Rewrites} & \textbf{Unrelated} & $\boldsymbol{M}$ & \textbf{Rewrites} & \textbf{Unrelated} & $\boldsymbol{M}$ \\
\midrule
\emph{Deep-Live-Cam}    & 61.8 & 20.0 & $+41.8$ & 45.7 & 11.3 & $+34.4$ \\
\emph{Algo}             & 77.8 &  0.0 & $+77.8$ & 56.2 & 10.9 & $+45.3$ \\
\emph{AgenticSeek}      & 37.0 &  7.0 & $+30.0$ & 46.0 & 13.1 & $+32.9$ \\
\emph{ChatTTS}          & 32.4 &  7.3 & $+25.1$ & 43.5 & 10.2 & $+33.2$ \\
\emph{PDFMathTranslate} & 53.5 & 12.1 & $+41.4$ & 37.3 & 15.2 & $+22.1$ \\
\emph{Heretic}          & 59.3 & 12.7 & $+46.6$ & 57.7 & 15.5 & $+42.1$ \\
\midrule
All                     & 53.6 &  9.8 & $+43.8$ & 47.7 & 12.7 & $+35.0$ \\
\bottomrule
\end{tabular}
\end{table}

\begin{table}[t]
\caption{RQ3. Retention (\%) of the original's fingerprint in the control and in the nine rewrites (mean), for the atomic and call-graph layers. $\Delta$ is $\Delta_{\mathrm{STIC}}$ in percentage points, and $\Delta{>}0$ counts the rewrites that retain more than the control.}
\label{tab:rq3-stic-ret}
\small
\setlength{\tabcolsep}{4pt}
\begin{tabular}{@{}lrrrrrrrr@{}}
\toprule
 & \multicolumn{4}{c}{\textbf{Atomic}} & \multicolumn{4}{c}{\textbf{Call graph}} \\
\cmidrule(lr){2-5}\cmidrule(l){6-9}
\textbf{Project} & \textbf{Control} & \textbf{Rewrites} & $\boldsymbol{\Delta}$ & $\boldsymbol{\Delta{>}0}$ & \textbf{Control} & \textbf{Rewrites} & $\boldsymbol{\Delta}$ & $\boldsymbol{\Delta{>}0}$ \\
\midrule
\emph{Deep-Live-Cam}    & 44.0 & 41.5 & $-2.5$  & 4/9 & 31.5 & 30.5 & $-0.9$  & 4/9 \\
\emph{Algo}             & 17.8 & 49.3 & $+31.5$ & 9/9 &  8.6 & 32.4 & $+23.8$ & 8/9 \\
\emph{AgenticSeek}      & 71.1 & 42.4 & $-28.7$ & 0/9 & 60.3 & 36.4 & $-23.9$ & 1/9 \\
\emph{ChatTTS}          & 42.9 & 49.0 & $+6.2$  & 5/9 & 32.0 & 41.6 & $+9.6$  & 8/9 \\
\emph{PDFMathTranslate} & 48.0 & 31.5 & $-16.5$ & 0/9 & 34.6 & 24.1 & $-10.5$ & 0/9 \\
\emph{Heretic}          & 74.1 & 53.2 & $-20.9$ & 1/9 & 47.4 & 41.3 & $-6.1$  & 3/9 \\
\midrule
All                     & 49.7 & 44.5 & $-5.2$  & 19/54 & 35.7 & 34.4 & $-1.3$ & 24/54 \\
\bottomrule
\end{tabular}
\end{table}

\find{{\bf [RQ-3] \ding{42}}
Import and fingerprint retention separate rewrites from unrelated projects by 44 and 35 points, but not from independent implementations of the same task.
Controls retain as much of the original as rewrites do in four of six projects, and only 19 of 54 rewrites exceed their control.
Retention rises with token similarity, so it adds little where similarity is low.
These results rest on one control per project, extended after the campaign.
}
\section{Discussion}
\label{sec:discussion}
We first interpret the three results and then draw their consequences for maintainers, for detector research and for providers of coding agents.

\subsection{Interpretation}

\paragraph{Low scores.}
A low token score shows that two trees differ on the surface.
Our rewrites reach such scores with the original in the workspace, 42 of 54 below 25\% and six below 5\%.
A low score therefore does not show that the new code was written without the original.
The converse does not hold either.
Rewrites of different projects score up to 10.5\% against each other, and 22 of our 54 rewrites score no higher than that against their own original.
For these 22, JPlag reports the level it reports for unrelated code.
The 1.29\% cited in the \texttt{chardet} case lies in this range, so the score is compatible with both an independent implementation and a rewrite made with the original at hand.

This limit concerns ground-up rewrites.
Under the layout-preserving and paradigm-shift prompts of RQ1, median similarity stays at 64\% and 51\%, and JPlag does what it was built for.
The reading of a token score is thus asymmetric.
A high score is evidence of copying, and a low score is evidence of neither copying nor independence.
This follows from the design of the tools (Section~\ref{sec:background}).
They target textual clones, and a ground-up rewrite that keeps behaviour is a Type-4 clone of its original.

\paragraph{Sources of variation.}
The choice of agent explains almost none of the variance in similarity (0.1\%) and the project little (14\%).
Most of it lies between runs of the same configuration.
Several mechanisms contribute.
The agent settles on a different architecture at each run, and the rewrites of one configuration differ in size by a factor of up to 4.6.
Larger rewrites tend to score higher, although the association is weak (rank correlation 0.35 over the 54 runs).
Files of the original sometimes survive the cleanup, as in the GPT run of RQ1 and the \emph{Heretic} run at 80.4\%.
The agent sometimes restructures the code again during repair, as on \emph{Algo}.
The last two belong to what the protocol delivers, and Section~\ref{sec:threats} shows that the variation remains without them.

This variation gives a person who publishes a rewrite a choice.
They can run the agent several times and keep the lowest score.
In our data, three runs are enough.
In all 18 configurations at least one of the three runs scores below 25\%, and the lowest score of a configuration never exceeds 17.2\%.
In 16 of the 18, at least one run is below 25\% and also passes the harness.
The two exceptions are the \emph{ChatTTS} configurations in which no run passes.
A published score may thus be the minimum of several draws, and a reader has no means to know it.

\paragraph{Task-imposed facts.}
The facts that distinguish a rewrite from unrelated projects are mostly facts that the task imposes.
A face-swapping tool imports OpenCV, InsightFace and ONNX Runtime whoever writes it, and it names the same models and providers in its configuration.
Our fingerprint selects facts that are rare among six projects from different domains, so it selects such facts first.
This explains why the controls retain 43\% to 74\% of the atomic fingerprint in five of six projects without any contact with the original code.
It also explains why the margin over unrelated projects is large and uninformative.

In four projects the control retains more than the average rewrite.
One possible reason is that the two were produced under opposite instructions.
The agent was asked to rearchitect and moved away from the names and constants of the original, whereas the author of the control followed the conventions of the domain.
The rewrites that moved furthest are those with the lowest token similarity, which is where retention is lowest too.

Copyright practice knows this distinction between task and origin.
The \emph{Altai} test removes the elements dictated by efficiency or by external constraints before two programs are compared~\cite{altai1992}.
A provenance signal validated only against unrelated code skips this step and measures the task.
Our controls show that the step is needed.
They do not show that no signal exists, since rewrites of \emph{Algo} and \emph{ChatTTS} retain more than their control.

\subsection{Implications}

\paragraph{Maintainers and auditors.}
A single low similarity score should not be accepted as evidence of independence.
When one is offered, the number of runs and the rule used to select the published one should be requested with it.
In our study, neither token similarity nor retention revealed that the agent had the original.
Only the record of the session does, that is, the prompts, the logs and the content of the workspace.
Evidence about the process is therefore worth more than evidence about the output.

This brings the question back to clean-room practice.
Its value never lay in the dissimilarity of the result but in a documented separation between those who read the original and those who wrote the new code.
With agents, the separation has to be enforced on the workspace and recorded.
It cannot be enforced on the training data of the model, which is a limit that a clean-room claim based on an agent should state.

\paragraph{Detector evaluation.}
A detector for rewrites should be evaluated against independent implementations of the same task.
Unrelated projects are a necessary reference and an insufficient one, as our margins of 44 and 35 points show.
Scores should also be reported over repeated generations of the same rewrite, since one generation says little about the next.
Scores between unrelated repositories give the floor below which a threshold has no meaning, 10.5\% for JPlag on our corpus.
Such evaluations need more than one control per project.
With six negatives, our own AUC estimates have intervals too wide to conclude.
The 69 rewrites and six controls that we release are a starting point for a larger benchmark.

\paragraph{Agent providers.}
None of our prompts asks to evade a licence.
A request to rewrite a project with a new architecture is ordinary, so a refusal policy based on stated intent does not apply to it.
What a provider can offer is the session record, which is the only evidence that settled the question of access in our study.

\paragraph{Future work.}
Three extensions would strengthen these results.
Agent runs from a specification alone would show what similarity and retention an agent reaches without the original, and several controls per project by different authors would replace the single point we compare with.
Held-out inputs and differential testing would make a harness pass more informative than agreement on five or six cases.
Other languages, larger repositories and cross-language rewriting would show how far the findings extend.

\section{Threats to Validity}
\label{sec:threats}

\paragraph{Construct validity.}
Token similarity measures surface correspondence and says nothing about derivation in the legal sense.
We use the mean of the two directional JPlag similarities.
With the maximum, 35 rewrites instead of 42 fall below 25\%, and the instability is at least as large (repetition accounts for 57\% of the variance, and 10 of 18 configurations straddle $\tau$).
The threshold $\tau$ is operational, and all threshold results are given at three values.

A harness pass means agreement on five or six cases that execute 20\% to 58\% of the lines of the original.
Coverage of the rewrites themselves ranges from 5\% to 84\%, so parts of some passing rewrites are never executed.
The feasibility claim of RQ1 is limited to what the harness exercises.
Cost is measured in session time and dialogue tokens, and we report no monetary cost.

The fingerprint is selected over six originals and would differ with another corpus.
The five unrelated projects come from other domains, which makes the margin $M$ an easy reference to exceed.

\paragraph{Internal validity.}
Two features of the protocol affect similarity.
Because the agent edits in place, files of the original can remain in a rewrite.
Without \emph{Heretic}, where this occurred, repetition still accounts for the largest share of the variance (52\%) and 4 of 15 configurations straddle $\tau$.
Repair went beyond integration changes on \emph{Algo}.
On scores taken before repair, repetition accounts for 54\% of the variance and 9 of 18 configurations straddle $\tau$.
The instability, therefore, does not depend on either feature.
Both show, however, that the agent did not always follow the constraints of the prompts, and that our check on archived trees did not catch every case.

The controls are the weakest part of the design.
There is one per project, written by a single author who had read the originals, wrote the capability specifications and the behavioural test suites from that knowledge, and then implemented and later extended each control under the 0\% JPlag acceptance check after the first retention results.

On \emph{Deep-Live-Cam} the sign of $\Delta_{\mathrm{STIC}}$ depends on this extension, and the \emph{Algo} control has a narrower scope than the original.
The controls were also required to score 0\% under JPlag, a condition that an independent implementation need not meet and that may have pushed their author away from natural choices.
RQ3 is a sensitivity result and not a test planned in advance.

One operator ran all sessions. A single operator ran every cloning and repair session in the campaign. That operator's only inputs were the fixed prompts and \texttt{Continue} messages. This limits but does not remove operator effects.
The operator's only inputs were fixed prompts and \texttt{Continue} messages, which limits but does not remove operator effects.

The models may have seen these public projects during training, and we cannot separate what an agent takes from the workspace from what the model recalls.
This does not affect our claims, since the setting gives the agent the original in any case.
Providers also update their models, so the same run may give other results at a later date.

\paragraph{Conclusion validity.}
With three repetitions per configuration, the variance decomposition is descriptive and we attach no significance test to its shares.
The absence of a detectable effect of the agent or the project does not establish that there is none.
RQ1 has one run per cell and supports no comparison between prompts or agents.
The AUC of RQ3 rests on six negatives, and its intervals are wide.

\paragraph{External validity.}
The corpus has six Python projects below 15{,}000 SLOC, selected purposively, and RQ1 uses one of them.
Five of the six are applications built on machine-learning libraries, in which much of the code calls third-party APIs.
Results may differ for projects whose value lies in their own algorithms, of which \texttt{chardet} is an example.
All runs use one agent tool, five models and three prompts.
We study rewriting in the same language with the original available.
We have no runs from a specification alone, so we cannot say what an agent without the original would score.

\section{Related Work}
\label{sec:related}

Three lines of work come close to this study, and each leaves open at least one of the quantities we measure.

\paragraph{Detectors under automated obfuscation.}
Plagiarism detectors have been tested against automated and LLM-based obfuscation.
Sa{\u{g}}lam et al.\ normalise token sequences so that JPlag withstands the insertion and reordering of statements~\cite{saglam2024tokennorm}, and Sa{\u{g}}lam and Schmid evaluate such defences on more than four million comparisons of student programs, including programs obfuscated or generated by GPT-4~\cite{saglam2025aiplagiarism}.
Park et al.\ detect LLM paraphrases of human-written code from coding-style features~\cite{park2025paraphrase}.
On the clone-detection side, GPTCloneBench provides semantic and cross-language clone pairs generated with GPT-3~\cite{alam2023gptclonebench}, and on this benchmark and BigCloneBench~\cite{svajlenko2015bigclonebench} Zhang and Saber report that GPT-3.5 and GPT-4 detect few Type-4 clones, with recall of 0.07 and 0.23~\cite{zhang2024clonellm}.
These studies work on functions or course assignments and score each transformed program once.
They do not cover repositories rewritten by an agent, and they do not ask whether a second rewrite of the same program would receive the same score.

\paragraph{Memorisation and licence compliance.}
A second line asks whether code models reproduce protected code.
Models are known to emit memorised training code~\cite{alkaswan2024traces}, CodeIPPrompt measures how close generated code comes to licensed code~\cite{yu2023codeipprompt}, and LiCoEval tests whether a model gives accurate licence information when its output is strikingly similar to an existing open-source implementation~\cite{licoeval2024}.
The closest study to ours is that of Colombo et al., who generated more than 70,000 Java method bodies with GPT-4 from the signatures and JavaDoc of copyleft methods and scored them with JPlag~\cite{colombo2025copyleft}.
They found that a larger class context increases the likelihood of reproducing copyleft code, that a higher temperature reduces it, and that an explicit request to avoid copying has no reliable effect.
Meeus et al.\ show that models also reproduce the functional logic of training code in textually dissimilar form, which they call functional memorisation and which overlap metrics do not detect~\cite{meeus2026functional}.
All of this work concerns single methods or functions completed from a signature, where similarity comes from what the model retained during training.
In our setting the original is handed to the agent, the unit is the repository, and the question is what a detector reports afterwards.

\paragraph{Repository construction.}
A third line evaluates whether agents can build a repository without its source.
NL2Repo-Bench gives the agent a requirements document and an empty workspace and runs the upstream tests on the result~\cite{nl2repo2025}.
RepoZero supplies API specifications and asks for a reimplementation in another language~\cite{repozero2026}, and ProgramBench supplies an executable with its documentation and accepts any implementation language~\cite{programbench2026}.
These benchmarks withhold the source by design and score behaviour only, so they never compare the result with the original and say nothing about rewriting when the source is available.

\paragraph{Other similarity measures.}
Measures based on tree edits~\cite{falleri2014gumtree,song2024tsed} could complement token matching when tokens diverge.
We keep to token-based detectors because they are the evidence produced in practice, and we report five further detectors in the appendix.

Our study differs from these lines on four points.
It works at repository scale in a single language, it gives the agent the original, it repeats each configuration, and it compares rewrites with independent implementations of the same task.
\section{Conclusion}
\label{sec:conclusion}

Token-similarity scores are offered as evidence that agent-made rewrites of copyleft projects are independent of their originals.
We examined what such scores can support on 69 rewrites of six Python projects, all produced with the original in the agent's workspace and with no human edit to the code.
The rewrites are cheap to obtain and most score low, 42 of 54 below 25\% under a ground-up prompt.
The score does not reproduce, however.
Repeating a configuration accounts for 61\% of its variance, and three runs suffice to obtain a score below 25\% in every configuration.
Import and fingerprint retention separate rewrites from unrelated projects but not from independent implementations of the same task.
A low similarity score is therefore weak evidence of independence, and a claim of independence should rest on records of how the code was produced.
Detectors for such rewrites should be validated against same-task controls and over repeated generations.

\section{Ethical Considerations}
\label{sec:ethics}

\paragraph{Dual use.}
This paper describes a protocol that produces low-similarity rewrites of copyleft code, and it could be used for that purpose.
We judge this risk limited.
The protocol uses ordinary prompts in a commodity tool and is already practised in public, as the \texttt{chardet} case shows.
What the paper adds is evidence on what similarity scores can support, which serves those who must assess such rewrites.

\paragraph{Licences.}
The six originals are under AGPL-3.0 or GPL-3.0, which permit the modification and redistribution carried out in this study.
We do not claim that the rewrites are free of these licences, and we distribute them under the licence of their original.

\paragraph{Sensitive projects.}
Two of the projects are a face-swapping tool and a tool that removes refusals from language models.
Our rewrites reproduce functionality that is already public and add no capability.
The harnesses use synthetic or author-controlled inputs and no personal data.

\paragraph{Participants.}
The study involved no human participants.

\clearpage
\bibliographystyle{ACM-Reference-Format}
\bibliography{acmart}

@misc{chardet327,
  author       = {Pilgrim, Mark},
  title        = {No right to relicense this project},
  howpublished = {GitHub issue \#327, chardet/chardet},
  year         = {2026},
  url          = {https://github.com/chardet/chardet/issues/327}
}

@misc{willison2026chardet,
  author = {Willison, Simon},
  title  = {Can coding agents relicense open source through a ``clean room'' implementation of code?},
  year   = {2026},
  month  = mar,
  url    = {https://simonwillison.net/2026/Mar/5/chardet/}
}

@article{lwn2026chardet,
  author = {Corbet, Jonathan},
  title = {The Relicensing of {chardet}},
  journal = {{LWN.net}},
  year = {2026},
  url = {https://lwn.net/Articles/1061534/},
  note = {Accessed 31 July 2026}
}

@article{roy2009comparison,
title = {Comparison and evaluation of code clone detection techniques and tools: A qualitative approach},
journal = {Science of Computer Programming},
volume = {74},
number = {7},
pages = {470-495},
year = {2009},
issn = {0167-6423},
doi = {https://doi.org/10.1016/j.scico.2009.02.007},
url = {https://www.sciencedirect.com/science/article/pii/S0167642309000367},
author = {Chanchal K. Roy and James R. Cordy and Rainer Koschke}
}

@INPROCEEDINGS{dipenta2010licensing,
  author={Di Penta, Massimiliano and German, Daniel M. and Guéhéneuc, Yann-Gaël and Antoniol, Giuliano},
  booktitle={2010 ACM/IEEE 32nd International Conference on Software Engineering}, 
  title={An exploratory study of the evolution of software licensing}, 
  year={2010},
  volume={1},
  number={},
  pages={145-154},
  doi={10.1145/1806799.1806824}}

@article{vendome2017license,
author = {Vendome, Christopher and Bavota, Gabriele and Penta, Massimiliano Di and Linares-V{\'a}squez, Mario and German, Daniel and Poshyvanyk, Denys},
title = {License usage and changes: a large-scale study on gitHub},
year = {2017},
issue_date = {June      2017},
publisher = {Kluwer Academic Publishers},
address = {USA},
volume = {22},
number = {3},
issn = {1382-3256},
url = {https://doi.org/10.1007/s10664-016-9438-4},
doi = {10.1007/s10664-016-9438-4},
journal = {Empirical Softw. Engg.},
month = jun,
pages = {1537–1577},
numpages = {41}
}

@inproceedings{
jimenez2024swebench,
title={{SWE}-bench: Can Language Models Resolve Real-world Github Issues?},
author={Carlos E Jimenez and John Yang and Alexander Wettig and Shunyu Yao and Kexin Pei and Ofir Press and Karthik R Narasimhan},
booktitle={The Twelfth International Conference on Learning Representations},
year={2024},
url={https://openreview.net/forum?id=VTF8yNQM66}
}

@misc{altai1992,
  author       = {{U.S. Court of Appeals for the Second Circuit}},
  title        = {{Computer Associates International, Inc. v. Altai, Inc.}},
  howpublished = {982 F.2d 693 (2d Cir. 1992)},
  year         = {1992},
  note         = {Abstraction--filtration--comparison test for software copyright}
}

@InProceedings{yu2023codeipprompt,
  title = 	 {{C}ode{IPP}rompt: Intellectual Property Infringement Assessment of Code Language Models},
  author =       {Yu, Zhiyuan and Wu, Yuhao and Zhang, Ning and Wang, Chenguang and Vorobeychik, Yevgeniy and Xiao, Chaowei},
  booktitle = 	 {Proceedings of the 40th International Conference on Machine Learning},
  pages = 	 {40373--40389},
  year = 	 {2023},
  editor = 	 {Krause, Andreas and Brunskill, Emma and Cho, Kyunghyun and Engelhardt, Barbara and Sabato, Sivan and Scarlett, Jonathan},
  volume = 	 {202},
  series = 	 {Proceedings of Machine Learning Research},
  month = 	 {23--29 Jul},
  publisher =    {PMLR},
  url = 	 {https://proceedings.mlr.press/v202/yu23g.html}
}

@inproceedings{licoeval2024,
author = {Xu, Weiwei and Gao, Kai and He, Hao and Zhou, Minghui},
title = {LiCoEval: Evaluating LLMs on License Compliance in Code Generation},
year = {2025},
isbn = {9798331505691},
publisher = {IEEE Press},
url = {https://doi-org.proxy.bnl.lu/10.1109/ICSE55347.2025.00052},
doi = {10.1109/ICSE55347.2025.00052},
booktitle = {Proceedings of the IEEE/ACM 47th International Conference on Software Engineering},
pages = {1665–1677},
numpages = {13},
location = {Ottawa, Ontario, Canada},
series = {ICSE '25}
}

@ARTICLE{bellon2007clone,
  author={Bellon, Stefan and Koschke, Rainer and Antoniol, Giulio and Krinke, Jens and Merlo, Ettore},
  journal={IEEE Transactions on Software Engineering}, 
  title={Comparison and Evaluation of Clone Detection Tools}, 
  year={2007},
  volume={33},
  number={9},
  pages={577-591},
  doi={10.1109/TSE.2007.70725}}

@article{prechelt2002jplag,
  author = {Prechelt, Lutz and Malpohl, Guido and Philippsen, Michael},
  title = {Finding Plagiarisms among a Set of Programs with {JPlag}},
  journal = {Journal of Universal Computer Science},
  volume = {8},
  number = {11},
  pages = {1016--1038},
  year = {2002},
  doi = {10.3217/jucs-008-11-1016},
  publisher = {J.UCS Consortium}
}

@inproceedings{schleimer2003,
author = {Schleimer, Saul and Wilkerson, Daniel S. and Aiken, Alex},
title = {Winnowing: local algorithms for document fingerprinting},
year = {2003},
isbn = {158113634X},
publisher = {Association for Computing Machinery},
address = {New York, NY, USA},
url = {https://doi.org/10.1145/872757.872770},
doi = {10.1145/872757.872770},
booktitle = {Proceedings of the 2003 ACM SIGMOD International Conference on Management of Data},
pages = {76–85},
numpages = {10},
location = {San Diego, California},
series = {SIGMOD '03}
}

@inproceedings{falleri2014gumtree,
  author = {Falleri, Jean-R{\'e}my and Morandat, Flor{\'e}al and Blanc, Xavier and Martinez, Matias and Monperrus, Martin},
  title = {Fine-grained and Accurate Source Code Differencing},
  booktitle = {Proceedings of the 29th {ACM}/{IEEE} International Conference on Automated Software Engineering ({ASE})},
  pages = {313--324},
  year = {2014},
  publisher = {{ACM}},
  doi = {10.1145/2642937.2642982}
}

@inproceedings{song2024tsed,
    title = "Revisiting Code Similarity Evaluation with Abstract Syntax Tree Edit Distance",
    author = "Song, Yewei  and
      Lothritz, Cedric  and
      Tang, Daniel  and
      Bissyand{\'e}, Tegawend{\'e}  and
      Klein, Jacques",
    editor = "Ku, Lun-Wei  and
      Martins, Andre  and
      Srikumar, Vivek",
    booktitle = "Proceedings of the 62nd Annual Meeting of the Association for Computational Linguistics (Volume 2: Short Papers)",
    month = aug,
    year = "2024",
    address = "Bangkok, Thailand",
    publisher = "Association for Computational Linguistics",
    url = "https://aclanthology.org/2024.acl-short.3/",
    doi = "10.18653/v1/2024.acl-short.3",
    pages = "38--46"
}

@INPROCEEDINGS{zhang2024clonellm,
  author={Zhang, Zixian and Saber, Takfarinas},
  booktitle={2024 4th International Conference on Code Quality (ICCQ)}, 
  title={Assessing the Code Clone Detection Capability of Large Language Models}, 
  year={2024},
  volume={},
  number={},
  pages={75-83},
  doi={10.1109/ICCQ60895.2024.10576803}}

@INPROCEEDINGS{alam2023gptclonebench,
  author={Alam, Ajmain I. and Roy, Palash R. and Al-Omari, Farouq and Roy, Chanchal K. and Roy, Banani and Schneider, Kevin A.},
  booktitle={2023 IEEE International Conference on Software Maintenance and Evolution (ICSME)}, 
  title={GPTCloneBench: A comprehensive benchmark of semantic clones and cross-language clones using GPT-3 model and SemanticCloneBench}, 
  year={2023},
  volume={},
  number={},
  pages={1-13},
  doi={10.1109/ICSME58846.2023.00013}}

@INPROCEEDINGS{svajlenko2015bigclonebench,
  author={Svajlenko, Jeffrey and Roy, Chanchal K.},
  booktitle={2015 IEEE International Conference on Software Maintenance and Evolution (ICSME)}, 
  title={Evaluating clone detection tools with BigCloneBench}, 
  year={2015},
  volume={},
  number={},
  pages={131-140},
  doi={10.1109/ICSM.2015.7332459}}

@inproceedings{colombo2025copyleft,
author = {Colombo, Gaia and Mariani, Leonardo and Micucci, Daniela and Riganelli, Oliviero},
title = {On the Possibility of Breaking Copyleft Licenses When Reusing Code Generated by ChatGPT},
year = {2025},
publisher = {IEEE Press},
url = {https://doi.org/10.1109/ICPC66645.2025.00060},
doi = {10.1109/ICPC66645.2025.00060},
booktitle = {2025 IEEE/ACM 33rd International Conference on Program Comprehension (ICPC)},
pages = {1–12},
numpages = {12},
location = {Ottawa, ON, Canada}
}

@misc{repozero2026,
      title={RepoZero: Can LLMs Generate a Code Repository from Scratch?}, 
      author={Zhaoxi Zhang and Yiming Xu and Jiahui Liang and Weikang Li and Xiaoshuai Chen and Liwei Qian and Xin Pei and Jizhou Huang and Run Sun and Yunfang Wu},
      year={2026},
      eprint={2605.07122},
      archivePrefix={arXiv},
      primaryClass={cs.SE},
      url={https://arxiv.org/abs/2605.07122}, 
}

@misc{programbench2026,
  title = {{ProgramBench}: Can Language Models Rebuild Programs From Scratch?},
  author = {Yang, John and Lieret, Kilian and Ma, Jeffrey and Thakkar, Parth and Pedchenko, Dmitrii and Sootla, Sten and McMilin, Emily and Yin, Pengcheng and Hou, Rui and Synnaeve, Gabriel and Yang, Diyi and Press, Ofir},
  year = {2026},
  eprint = {2605.03546},
  archivePrefix = {arXiv},
  primaryClass = {cs.SE},
  url = {https://arxiv.org/abs/2605.03546}
}

@inproceedings{yang2024sweagent,
author = {Yang, John and Jimenez, Carlos E. and Wettig, Alexander and Lieret, Kilian and Yao, Shunyu and Narasimhan, Karthik and Press, Ofir},
title = {SWE-agent: agent-computer interfaces enable automated software engineering},
year = {2024},
isbn = {9798331314385},
publisher = {Curran Associates Inc.},
address = {Red Hook, NY, USA},
booktitle = {Proceedings of the 38th International Conference on Neural Information Processing Systems},
articleno = {1601},
numpages = {125},
location = {Vancouver, BC, Canada},
series = {NIPS '24}
}

@inproceedings{
  wang2024openhands,
  title={OpenHands: An Open Platform for {AI} Software Developers as Generalist Agents},
  author={Xingyao Wang and Boxuan Li and Yufan Song and Frank F. Xu and Xiangru Tang and Mingchen Zhuge and Jiayi Pan and Yueqi Song and Bowen Li and Jaskirat Singh and Hoang H. Tran and Fuqiang Li and Ren Ma and Mingzhang Zheng and Bill Qian and Yanjun Shao and Niklas Muennighoff and Yizhe Zhang and Binyuan Hui and Junyang Lin and Robert Brennan and Hao Peng and Heng Ji and Graham Neubig},
  booktitle={The Thirteenth International Conference on Learning Representations},
  year={2025},
  url={https://openreview.net/forum?id=OJd3ayDDoF}
}

@inproceedings{alkaswan2024traces,
author = {Al-Kaswan, Ali and Izadi, Maliheh and van Deursen, Arie},
title = {Traces of Memorisation in Large Language Models for Code},
year = {2024},
isbn = {9798400702174},
publisher = {Association for Computing Machinery},
address = {New York, NY, USA},
url = {https://doi.org/10.1145/3597503.3639133},
doi = {10.1145/3597503.3639133},
booktitle = {Proceedings of the IEEE/ACM 46th International Conference on Software Engineering},
articleno = {78},
numpages = {12},
location = {Lisbon, Portugal},
series = {ICSE '24}
}

@inproceedings{saglam2024tokennorm,
author = {Sa{\u g}lam, Timur and Br{\"o}del, Moritz and Schmid, Larissa and Hahner, Sebastian},
title = {Detecting Automatic Software Plagiarism via Token Sequence Normalization},
year = {2024},
isbn = {9798400702174},
publisher = {Association for Computing Machinery},
address = {New York, NY, USA},
url = {https://doi.org/10.1145/3597503.3639192},
doi = {10.1145/3597503.3639192},
booktitle = {Proceedings of the IEEE/ACM 46th International Conference on Software Engineering},
articleno = {113},
numpages = {13},
location = {Lisbon, Portugal},
series = {ICSE '24}
}

@misc{saglam2025aiplagiarism,
      title={Evaluating Software Plagiarism Detection in the Age of AI: Automated Obfuscation and Lessons for Academic Integrity}, 
      author={Timur Sağlam and Larissa Schmid},
      year={2025},
      eprint={2505.20158},
      archivePrefix={arXiv},
      primaryClass={cs.SE},
      url={https://arxiv.org/abs/2505.20158}, 
}

@misc{park2025paraphrase,
      title={Detection of LLM-Paraphrased Code and Identification of the Responsible LLM Using Coding Style Features}, 
      author={Shinwoo Park and Hyundong Jin and Jeong-won Cha and Yo-Sub Han},
      year={2026},
      eprint={2502.17749},
      archivePrefix={arXiv},
      primaryClass={cs.AI},
      url={https://arxiv.org/abs/2502.17749}, 
}

@misc{meeus2026functional,
      title={Detecting Functional Memorization in Code Language Models}, 
      author={Matthieu Meeus and Anil Ramakrishna and Shengyuan Hu and Matthew Grange and Zheng Xu and Luca Melis},
      year={2026},
      eprint={2606.12764},
      archivePrefix={arXiv},
      primaryClass={cs.LG},
      url={https://arxiv.org/abs/2606.12764}, 
}

@misc{dong2025surveyagents,
      title={A Survey on Code Generation with LLM-based Agents}, 
      author={Yihong Dong and Xue Jiang and Jiaru Qian and Tian Wang and Kechi Zhang and Zhi Jin and Ge Li},
      year={2025},
      eprint={2508.00083},
      archivePrefix={arXiv},
      primaryClass={cs.SE},
      url={https://arxiv.org/abs/2508.00083}, 
}

@misc{nl2repo2025,
      title={NL2Repo-Bench: Towards Long-Horizon Repository Generation Evaluation of Coding Agents}, 
      author={Jingzhe Ding and Shengda Long and Changxin Pu and Huan Zhou and Hongwan Gao and Xiang Gao and Chao He and Yue Hou and Fei Hu and Zhaojian Li and Weiran Shi and Zaiyuan Wang and Daoguang Zan and Chenchen Zhang and Xiaoxu Zhang and Qizhi Chen and Xianfu Cheng and Bo Deng and Qingshui Gu and Kai Hua and Juntao Lin and Pai Liu and Mingchen Li and Xuanguang Pan and Zifan Peng and Yujia Qin and Yong Shan and Zhewen Tan and Weihao Xie and Zihan Wang and Yishuo Yuan and Jiayu Zhang and Enduo Zhao and Yunfei Zhao and He Zhu and Liya Zhu and Chenyang Zou and Ming Ding and Jianpeng Jiao and Jiaheng Liu and Minghao Liu and Qian Liu and Chongyang Tao and Jian Yang and Tong Yang and Zhaoxiang Zhang and Xinjie Chen and Wenhao Huang and Ge Zhang},
      year={2026},
      eprint={2512.12730},
      archivePrefix={arXiv},
      primaryClass={cs.CL},
      url={https://arxiv.org/abs/2512.12730}, 
}

@misc{gplv3,
  author = {{Free Software Foundation}},
  title = {{GNU} General Public License, Version~3},
  year = {2007},
  howpublished = {\url{https://www.gnu.org/licenses/gpl-3.0.html}},
}

@misc{agplv3,
  author = {{Free Software Foundation}},
  title = {{GNU} Affero General Public License, Version~3},
  year = {2007},
  howpublished = {\url{https://www.gnu.org/licenses/agpl-3.0.html}},
}

\end{document}